\documentclass[12pt]{iopart}
 
\usepackage{iopams}  

\usepackage[export]{adjustbox}

\usepackage{graphicx}
\usepackage{dcolumn}
\usepackage{bm}
\usepackage{makecell}
\usepackage{xcolor}
\usepackage{enumerate}
\usepackage[caption=false]{subfig}
\usepackage{float}
\usepackage[unicode=true,pdfusetitle,
 bookmarks=true,bookmarksnumbered=false,bookmarksopen=false,
 breaklinks=false,pdfborder={0 0 1},backref=false,colorlinks=true]
 {hyperref}
\hypersetup{linkcolor=blue,urlcolor=blue,citecolor=blue}

\newcommand{\be}{\begin{equation}}
\newcommand{\ee}{\end{equation}}
\newcommand{\bd}{\begin{displaymath}}
\newcommand{\ed}{\end{displaymath}}
\newcommand{\BE}{\begin{eqnarray}}
\newcommand{\EE}{\end{eqnarray}}

\newcommand{\bx}{\ensuremath{\mathbf{x}}}

\newcommand{\bn}{\ensuremath{\mathbf{n}}}

\newcommand{\avg}[1]{\left\langle{#1}\right\rangle}

\newcommand{\eqref}[1]{(\ref{#1})}

\begin{document}

\title{Demographic noise in complex ecological communities}

 \author{Ferran Larroya}
\ead{ferran.larroya@ub.edu}
\address{Departament de Física de la Matèria Condensada, Universitat de Barcelona, Martí i Franquès 1, E-08028, Barcelona, Spain.}
\address{Universitat de Barcelona Institute of Complex Systems, Martí i Franquès, 1, E-08028 Barcelona, Spain.}
\address{Instituto de F\'isica Interdisciplinar y Sistemas Complejos, IFISC (CSIC-UIB), Campus Universitat Illes Balears, E-07122 Palma de Mallorca, Spain}

\author{Tobias Galla}
\ead{tobias.galla@ifisc.uib-csic.es}
 
\address{Instituto de F\'isica Interdisciplinar y Sistemas Complejos, IFISC (CSIC-UIB), Campus Universitat Illes Balears, E-07122 Palma de Mallorca, Spain}

\begin{abstract}
We introduce an individual-based model of a complex ecological community with random interactions. The model contains a large number of species, each with a finite population of individuals, subject to discrete reproduction and death events. The interaction coefficients determining the rates of these events is chosen from an ensemble of random matrices, and is kept fixed in time. The set-up is such that the model reduces to the known generalised Lotka--Volterra equations with random interaction coefficients in the limit of an infinite population for each species. Demographic noise in the individual-based model means that species which would survive in the Lotka--Volterra model can become extinct. These noise-driven extinctions are the focus of the paper. We find that, for increasing complexity of interactions, ecological communities generally become less prone to extinctions induced by demographic noise.  An exception are systems composed entirely of predator-prey pairs. These systems are known to be stable in deterministic Lotka--Volterra models with random interactions, but, as we show, they are nevertheless particularly vulnerable to fluctuations. \end{abstract}

\section{Introduction}
Ecological communities are inherently complex systems, in which species interact with each other, generating intricate equilibria, patterns and dynamics. The factors that promote or hinder stability are of great interest, together with the question of how much complexity an ecosystem can sustain at stable equilibria. This is at the heart of the so-called `stability-diversity' or `stability-complexity' debate, started fifty years ago by Robert May \cite{may,maybook,mccann,allesina_review}. May started from a hypothetical equilibrium of an ecosystem with many species, and then modelled the Jacobian at this equilibrium as a random matrix (we will refer to ecosystems with random interactions as `complex'). He then used knowledge from random matrix theory (i.e., the eigenvalue spectra of random matrices) to conclude that an increasing number of species ought to make an ecological equilibrium unstable in random ecosystems. This lead him to speculate that there must be some `devious strategy' nature is using to maintain complex equilibria \cite{maybook}. May's model has been subject to a variety of criticism (for an overview see e.g. \cite{namba}), and has been extended and refined in a number of different ways. This includes for example generalisations to spatial systems or to more structured random matrices \cite{allesinatang2, gravel, baron2020dispersal}.

Much of this work does not specify an actual dynamics for the ecosystem. A parallel stream of work therefore focuses on generalised Lotka--Volterra equations or related differential equations, with a random interaction structure similar to the hypothetical Jacobian matrices in May's approach \cite{diederich1989replicators,opper1992phase, Yoshino_2007, Yoshino_PRE, bunin2016interaction, bunin2017, froy, Galla_2018, sidhomgalla, birolibunin, altieri2021}. The analysis of such systems is not straightforward, the nonlinearity of the dynamics combined with the random interactions make an analytical solution for individual instances impossible. The problem can however be recognised as analogous to questions in the theory of spin glasses and disordered systems, originally devised to describe certain materials at low temperatures \cite{mpv}. Analytical tools are available from statistical physics to address dynamical systems with quenched (i.e., time-independent) random interaction coefficients \cite{mpv, dedominicis1978dynamics,msr,hertz2016path}. These methods were first applied to random-replicator models in \cite{diederich1989replicators,opper1992phase}, these are systems that are structurally very similar to generalised Lotka--Volterra equations (gLVE). Subsequent work addresses gLVEs directly, and has established when such systems are stable or unstable respectively, and characterised the equilibria in the stable phase. This work is based on generating functionals \cite{Galla_2018} or the so-called cavity method \cite{bunin2016interaction,bunin2017,froy}. For systems in which the interactions between species are symmetric the celebrated replica method can also be applied \cite{birolibunin,altieri2021}.

The starting point of these existing analyses is set by the random replicator or generalised Lotka--Volterra dynamics. While the interaction coefficients are drawn at random at the beginning the subsequent dynamics is assumed to be deterministic. Analytical work then focuses on describing where in parameter space stable fixed points are found, and what the statistics of these fixed points are. Crucially though, the effects of so-called demographic (or intrinsic) noise are often left out. The deterministic gLVE describe continuous abundances of species, and discard the inherent randomness of birth and death events. Neglecting this stochasticity is valid when the number of individuals for each species is large, any death or birth then only changes abundances by a small amount, and the stochasticity in the birth and death events `averages out'. However, complex interaction models are also used to describe microbial populations \cite{coyte, yonatan} and in game theory \cite{gokhale, gallafarmer}, and it is known that demographic noise can significantly alter the behaviour of finite evolving populations in these areas \cite{traulsenhauert, constable_PNAS}. In this work we therefore set out to study some of the effects of demographic (or intrinsic) noise on the dynamics of complex Lotka--Volterra dynamics. We note that the work in \cite{birolibunin,altieri2021} makes first steps towards including demographic noise in complex Lotka--Volterra systems. However intrinsic stochasticity is there modelled by adding a Gaussian noise term to the gLVE.

In the present work we take a different approach. We introduce an individual-based Lotka--Volterra system, describing a population in which there are finitely many discrete individuals for each species. These individuals undergo discrete birth and death events, with rates governed by their interactions. The model is set up such that there is a control parameter setting the typical number of individuals belonging to a species (the parameter can be thought of the area or volume in which the ecosystem resides). In this way we recover the known random gLVE by sending this parameter to infinity. Our work focuses specifically on the effects of intrinsic noise on the stable fixed points of the limiting deterministic system. We study how this  stochasticity can lead to additional extinctions and how the distribution of species abundances changes as a consequence. We ask what species in particular are at risk of such noise-driven extinctions. We note that individual-based models or other stochastic dynamics have been used for Lotka--Volterra dynamics for example in \cite{stochastic1,stochastic2,stochastic3,stochastic4,stochastic5,stochastic6,stochastic7,stochastic8,stochastic9}. With the exception of \cite{stochastic1} this previous literature does not focus on complex ecosystems however (that is ecosystems with random interaction coefficients). Our model contains two types of stochasticity. One is the quenched randomness of the interactions which chosen before the dynamics starts, but then remain fixed. The other type of noise describes randomness as the dynamics of the system unfolds. The combination of the two is where the main novelty of our work lies.

The remainder of the paper is organised as follows: In Sec.~\ref{sec:model} we briefly summarise the deterministic random gLVE equations, and then introduce the individual-based model and describe how it can be simulated. To set the scene, Sec.~\ref{sec:det} then contains a brief account of the known phenomenology of the deterministic gLVE model. This serves as a basis to compare the individual-based model against. Our main results are the contained in Sec.~\ref{sec:noise}. We investigate the effects of demographic noise and show how stochasticity changes species abundances relative to the deterministic model. Sec.~\ref{sec:concl} finally contains our conclusions and a discussion of our results.

\section{Model definitions}\label{sec:model}
In this section, we first summarise the conventional deterministic random Lotka--Volterra model, and then define an individual-based model of these dynamics.
\subsection{Deterministic generalised Lotka--Volterra dynamics with random interactions}
The model describes $N$ interacting species, labelled $i=1,...,N$. The abundance of species $i$ at time $t$ is denoted by $x_{i}(t)$, we collect the abundances into the vector $\bx=(x_1,\dots,x_N)$. 
The generalised Lotka--Volterra equations are then
\begin{equation}
\frac{dx_{i}(t)}{dt}=x_{i}(t)\left(K_{i}-x_i+\sum_{j\neq i} \alpha_{ij}x_{j}(t)\right).
\label{eq:lotkavolterra1}
\end{equation}
The coefficients $\alpha_{ij}$ describe the interactions between the species, representing the benefit or detriment that species $i$ receives from species $j$. Consequently, a negative coefficient $\alpha_{ij}$ indicates a detrimental effect of species $j$ on species $i$. If $\alpha_{ij}$ and  $\alpha_{ji}$ have opposite signs, then species $i$ and $j$ constitute a prey-predator pair (i.e., there is antagonistic interaction). If $\alpha_{ij}$ and $\alpha_{ji}$ are both negative, the two species compete with each other, and if both coefficients are positive then they have a mutually beneficial relation with one another. The quantity $K_{i}$ in Eq.~(\ref{eq:lotkavolterra1}) is the carrying capacity of  species $i$ in monoculture, this is the equilibrium value of the abundance of species $i$ if there is no interaction with any other species ($\alpha_{ij}=0$ $\forall j\neq i$).  In principle heterogeneous $K_i$ can be considered, however we focus on the simpler case $K_i\equiv 1$ for all $i$.

Following \cite{bunin2016interaction,bunin2017,birolibunin,Galla_2018} we choose the interaction coefficients from a fixed distribution at the beginning, they then remain constant as the dynamics proceeds. More precisely, the off-diagonal coefficients $\alpha_{ij}$ are drawn from a Gaussian distribution with the following  mean, variance and correlations,
\begin{equation}
 \overline{\alpha_{ij}}= \frac{\mu}{N}, \\
  \overline{\alpha_{ij}^{2}}-\frac{\mu^{2}}{N^{2}}= \frac{\sigma^{2}}{N}, \\
  \overline{\alpha_{ij}\alpha_{ji}}-\frac{\mu^{2}}{N^{2}}= \Gamma \frac{\sigma^{2}}{N},
 \label{eq:gaussian}
\end{equation}
where the overbar denotes an average over the ensemble of random interaction matrices. The scaling with $N$ is chosen to guarantee a well-defined thermodynamic limit $N\rightarrow \infty$ \cite{mpv}. In our notation the diagonal coefficients $\alpha_{ii}$ do not enter in Eq.~(\ref{eq:lotkavolterra1}), but we note the stabilising term $-x_i$ inside the bracket on the right-hand side, which could be absorbed in the interaction term by defining $\alpha_{ii}=-1$.

 A positive value of $\mu$ means that species tend to benefit from each others presence.  On the other hand, a negative $\mu$ indicates competition. The parameter $\sigma^{2}$ controls the variance of the interactions, and describes the degree of heterogeneity or complexity of the species interactions. Finally the symmetry parameter $\Gamma$ accounts for the correlations between the interaction coefficients of any pair of species, and can take values between $-1 \leq \Gamma \leq 1$.

In the limit $N\rightarrow \infty$, the fraction of predator-pray pairs in the system (i.e, pairs $i,j$ with $\alpha_{ij}\alpha_{ji}<0$) can be obtained performing a Gaussian integral over the two quadrants, $\lbrace \alpha_{ij}\geq 0, \alpha_{ji}\leq 0 \rbrace$ and $\lbrace \alpha_{ij}\leq 0, \alpha_{ji}\geq 0 \rbrace$, of the joint distribution of $\alpha_{ij}$ and $\alpha_{ji}$ (see e.g. \cite{gaussian}). This results in a non-linear and decreasing dependence of the fraction of predator-pray pairs $p$ on $\Gamma$ in the limit of large $N$, $p=\frac{1}{2}-\frac{1}{\pi}\sin^{-1}{\left( \Gamma \right)}$. In particular, for $\Gamma=0$, $\alpha_{ij}$ and $\alpha_{ji}$ are uncorrelated and there are $50\%$ of predator-prey pairs in the system. If $\Gamma=1$, then $\alpha_{ij}=\alpha_{ji}$ (fully correlated) and there are no predator-prey pairs. Finally, for $\Gamma=-1$ (full anti-correlation), all pairs in the system are of the predator-prey type.

Despite the fact that the interaction coefficients are chosen at random at the beginning, we will refer to in Eq.~(\ref{eq:lotkavolterra1}) as deterministic. This is because there is no further stochasticity during the actual dynamics.

\subsection{Individual-based model}
We now define an individual-based variant of the generalised random Lotka--Volterra dynamics. 
\subsubsection{Birth-death dynamics.}
In order to incorporate demographic stochasticity into the deterministic Lotka--Volterra model, we focus on a population of discrete individuals and consider the dynamics as a birth-death Markov process in continuous time, driven by discrete events. In each event, an individual of a species $i$ reproduces or dies, increasing or reducing respectively the abundance of that species. The rates for these events are $T_{i}^{+}$ and $T_{i}^{-}$ respectively. Therefore, at each time step there are $2N$ possible reactions or events. Similar approaches have been used in \cite{stochastic1, stochastic2, stochastic3} and especially in \cite{stochastic2020}.

We write $n_i(t)$ for the number of individuals of species $i$ at time $t$, and $\bn(t)=[n_1(t),\dots, n_N(t)]$. In order to define the rates, $T_i^\pm$, we first introduce a parameter $V$, characterising the area or the volume in which the population resides. We will chose the birth and death rates such that the deterministic gLVE model is recovered in the limit $V\to\infty$. We write $x_{i}(t)=n_{i}(t)/V$. The deterministic Lotka-Volterra equations (\ref{eq:lotkavolterra1}) (with $K_i\equiv 1$) can then be written as
\BE
\hspace{-2em}\frac{dn_{i}(t)}{dt} &=& n_{i}(t) - \frac{n_{i}^{2}(t)}{V} + \sum_{j\neq i,~ \alpha_{ij}>0} \alpha_{ij}\frac{n_{i}(t)n_{j}(t)}{V} - \sum_{j\neq i,~\alpha_{ij}<0
}\left|\alpha_{ij}\right|\frac{n_{i}(t)n_{j}(t)}{V} , 
\label{eq:lotkavolterrastochastic}
\EE
where we have separated the terms with positive $\alpha_{ij}$ from those with negative interaction coefficients. These terms set the rates at which individuals of species $i$ reproduce and die in the stochastic model. We define,
\BE
 T_{i}^{+}(\bn)&=&n_{i}(t)+\sum_{j\neq i,~\alpha_{ij}>0} \alpha_{ij}\frac{n_{i}(t)n_{j}(t)}{V}, \nonumber  \\
T_{i}^{-}(\bn)&=&\frac{n_{i}^{2}(t)}{V}+\sum_{j \neq i,~\alpha_{ij}<0} \left|\alpha_{ij}\right|\frac{n_{i}(t)n_{j}(t)}{V}.
\label{eq:rates}
\EE
There is therefore a baseline per capita reproduction rate of one [first term in $T_i^+(\bn)$], and reproduction rates are enhanced by positive interaction with other species. The first term in the death rate $T_i^-(\bn)$ describes intra-species competition, and the second term (increasing the death rate) represents detrimental effects on species $i$ by other species. Given that the $n_i$ scale linearly in $V$, the reaction rates are of order ${\cal O}(V)$, indicating that the total number of events in the system per unit time is proportional to $V$.

The stochastic process of $\bn$ defined by the rates $T_i^\pm(\bn)$ can formally be described by a master equation. Using standard methods \cite{vanKampen1992,gardiner1985handbook, Risken1996} one can then show that the moments $\avg{n_i}$ of the process follow the Lotka-Volterra dynamics in Eq.~(\ref{eq:lotkavolterrastochastic}) in the limit of large $V$ (in this limit, the distribution of $\bn$ at time $t$ becomes sharply peaked around its mean). In other words, the quantities $x_i=\lim_{V\to\infty} \avg{n_i}/V$ follow Eq.~(\ref{eq:lotkavolterra1}). 

We stress that the choice of rates $T_i^\pm(\bn)$ in Eq.~(\ref{eq:rates}) is not the only one leading to Eq.~(\ref{eq:lotkavolterra1}) in the deterministic limit. For example, one could consider combined birth-death events, in which an individual of one species is replaced by that of another. We do however expect the principal effects we observe in the stochastic model to be robust against variation of these details.

\subsubsection{Simulation method.}\label{sec:gillespie}
We use the well-known exact Gillespie algorithm to simulate the invididual-based model \cite{Gillespie1976,Gillespie1977}. This generates a statistically faithful ensemble of trajectories of the stochastic process. The main principles of the simulation are an exponentially distributed time increment until the next event, followed by a stochastic selection of the type of event. Both the parameter of the exponential and the probabilities with which the different possible events are chosen are set by the $T_i^\pm(\bn)$. In detail, the simulation proceeds as follows:
\begin{enumerate}
    \item[1.] Draw a random interaction matrix $\alpha_{ij}$ with the statistics in Eq.~(\ref{eq:gaussian}). Set the initial conditions for the abundances $n_{i}$ ($i=1,\dots,N$) at $t=0$.
    \item[2.] Calculate the rates $T_{i}^{\pm}(\bn)$, the total rate $W=\sum_i \{T_i^+(\bn)+T_i^-(\bn)\}$, and set $p_{i}^{\pm}=T_i^\pm(\bn)/W$.
    \item[3.] Draw a time increment $\tau$ from an exponential distribution with parameter $W$, and increment time $t\to t+\tau$.
    \item[4.] Execute one single reaction, chosen among all possible reactions with the probabilities $p_i^\pm$. (This event is a the birth of an individual of species $i$ with probability $p_i^+$, and a death event for $i$ with probability $p_i^-$.)
    \item[5.] Go to step 2, or stop the simulation if the designated end time is reached.
\end{enumerate}

\section{Background: behaviour of the deterministic model}\label{sec:det}
In this section we briefly describe the known phenomenology of the deterministic Lotka--Volterra system with random interactions in Eq.~(\ref{eq:lotkavolterra1}). This serves as a baseline for our study of effects induced by demographic noise in Sec.~\ref{sec:noise}.

\subsection{Generating functional analysis}
The gLVE system with random interaction matrices has been studied analytically in the limit $N\to\infty$ with tools from the theory of disordered systems, in particular using path-integral analyses \cite{Galla_2018,sidhomgalla} or the so-called cavity method \cite{bunin2016interaction, bunin2017,froy}. Static approaches to the symmetric system ($\Gamma=1$) use the celebrated replica method, see e.g. \cite{birolibunin,altieri2021}.

In the essence these approaches involve carrying out the disorder-average (often via a static or dynamic partition function), and then deriving an effective dynamics or an effective potential for a representative species. The fixed points of this dynamics (or the minima of the potential) can then be characterised further. Ultimately, these approaches all lead to the same results of course when they can be applied. However, generating-functional and cavity methods are more general in that they do not require any restrictions on $\Gamma$. The replica method on the other hand allows for a more in-depth analysis of the `energy' landscape of the model and of the resulting analogies to spin glasses \cite{birolibunin,altieri2021}.

We here do not describe the analysis in detail, instead we focus on the equations for the key order parameters describing the stable fixed points of the gLVE.

The gLVE system exhibits a stable phase, in which a given realisation of the random interaction matrix leads to one unique stable fixed point with overwhelming probability as $N\to \infty$. This phase is found at sufficiently low values of the paramaters $\mu$ and $\sigma$, see Sec.~\ref{sec:instab} for further details.

The key order parameters in this fixed point phase are the mean abundance per species at the fixed point, $M$, the second moment  of the fixed point abundances, $q$, and the so-called susceptibility $\chi$, which indicates how a fixed point is shifted in abundance space in response to a permanent external perturbation \cite{bunin2016interaction,bunin2017}.

From the path-integral analysis and a subsequent fixed-point ansatz (or alternatively, from the cavity approach), one derives the following relations for these order parameters \cite{bunin2016interaction, bunin2017, Galla_2018},
\BE
  \chi &=& \frac{1}{1-\Gamma \sigma^{2}\chi} w_0(\Delta), \nonumber \\
  M &=& \sqrt{q}\sigma \frac{1}{1-\Gamma \sigma^{2}\chi} w_1(\Delta), \nonumber  \\
  1 &=& \frac{\sigma^{2}}{(1-\Gamma \sigma^{2}\chi )^{2}} w_2(\Delta), \nonumber \\
\label{eq:orderparameters2}
\EE
with $w_\ell(\Delta)=\int_{-\infty}^\Delta \frac{dz}{\sqrt{2\pi}}e^{-z^{2}/2} (\Delta-z)^\ell$, for $\ell=0, 1, 2$, and $\Delta = \left( 1+\mu M \right) / \left( \sqrt{q}\sigma \right)$. 

From the solution of these equations, the species abundance distribution is found as
\be
P(x)=(1-\phi)\delta(x)+\phi G(x),
\ee
where $G(x)$ is a Gaussian distribution with mean $(1+\mu M)/(1-\Gamma\sigma^2\chi)$ and variance $q\sigma^2/(1-\Gamma\sigma^2\chi)^2$, clipped at $x=0$ (abundances only take non-negative values). The coefficient $\phi = w_0(\Delta)$ is the fraction of surviving species. Examples of these species abundance distributions are shown further below in Fig.~\ref{fig:sad} (a)--(c).

\subsection{Instabilities and phase diagram}\label{sec:instab}
\begin{figure}[t]
 \centering
 \includegraphics[width=0.6\textwidth]{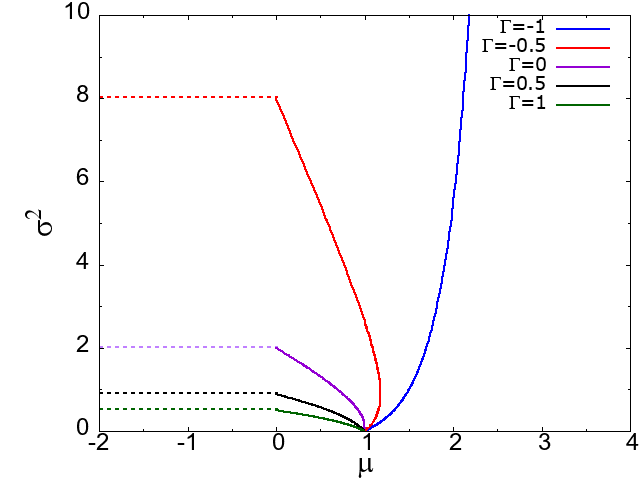}
 \caption{Phase diagram of the deterministic generalised Lotka-Volterra equations with random interaction matrices. The solid lines mark the onset of a diverging mean abundance per species ($M\to\infty$), the dashed lines are indicate where the linear instability sets in [Eq.~(\ref{eq:criticalsigma})]. The system is stable in the region below the dashed lines and to the left of the solid lines.}
 \label{fig:pg}
\end{figure}
Two types of instability of the fixed points in the gLVE system have been established \cite{bunin2016interaction,bunin2017,galla2018dynamically}. One is a linear instability, this sets in when
\begin{equation}
    \sigma^2=\sigma_c^{2}(\Gamma)=\frac{2}{(1+\Gamma)^{2}}.
    \label{eq:criticalsigma}
\end{equation}
Therefore for $\sigma^2 < \sigma_{c}^2$ (and in absence of the other type of instability) one expects the Lotka--Volterra dynamics to have stable fixed points in the limit of large number of species, $N$. The fixed points are unique, i.e. independent of initial conditions, for a given interaction matrix. For $\sigma^2 > \sigma_{c}^2$, the system can show several different types of behaviour. For example, the dynamics can remain volatile and potentially chaotic, or there can be a very large number of fixed points (initial conditions then determine which one of these is reached). For further details see for example \cite{sidhomgalla}.

The second type of instability occurs when the mean abundance per species, $M$, diverges. This indicates unbounded growth of the ecosystem. No closed-form expression is available for the point in parameter space where this happens, however the onset of this instability can be obtained numerically from Eqs. (\ref{eq:orderparameters2}), and imposing $1/M=0$ \cite{bunin2016interaction,bunin2017,Galla_2018}.

The resulting phase diagram in the $(\mu,\sigma^2)$-plane is shown for different values of $\Gamma$ in Fig.~\ref{fig:pg}. The system is stable and Eqs.~(\ref{eq:orderparameters2}) are valid in the region below the dashed lines (linear instability) and to the left of the solid lines (diverging abundance).
 
\section{Effects of demographic noise}\label{sec:noise}
We now study the effects of demographic noise on the ecosystem. Our analysis focuses on parameters in the stable phase of the gLVE. The rationale for this focus is discussed in the conclusions in Sec.~\ref{sec:concl}. 

\subsection{Effects on trajectories}
\begin{figure}[t]
 \centering
 \includegraphics[width=0.6\textwidth]{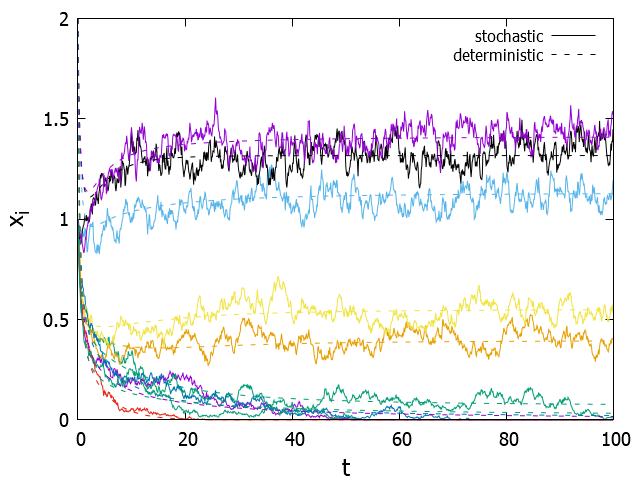}
 \caption{Sample trajectory of the stochastic system (solid wiggly lines, $V=500$) and of the deterministic system (dashed lines). The figure shows the abundaces of a selection of species as a function of time. Data is from one realisation of the interaction matrix with $N=50, \Gamma=1, \sigma^2=0.2, \mu=-1$.}
 \label{fig:traj}
\end{figure}

We start by illustrating trajectories of the deterministic and stochastic systems in Fig.~\ref{fig:traj}, for one particular realisation of the interaction matrix. Model parameters are chosen such that the deterministic system approaches a stable fixed point (indicated by the dashed lines). As seen in the figure the abundances of the stochastic system broadly fluctuate about the deterministic fixed-point values. By virtue of the central limit theorem the variance of these fluctuations can be expected to scale as $V^{-1}$. 

As discussed in the previous section, some species become extinct at the deterministic fixed point. Those species are normally also found to die out in the system with demographic noise. Additionally, there can be species which survive in the absence of noise, but which are driven to extinction by fluctuations in the stochastic system. An example can be seen in the bottom right of Fig.~\ref{fig:traj}. 

\begin{figure*}[t]
 \centering
 \includegraphics[width=0.85\textwidth]{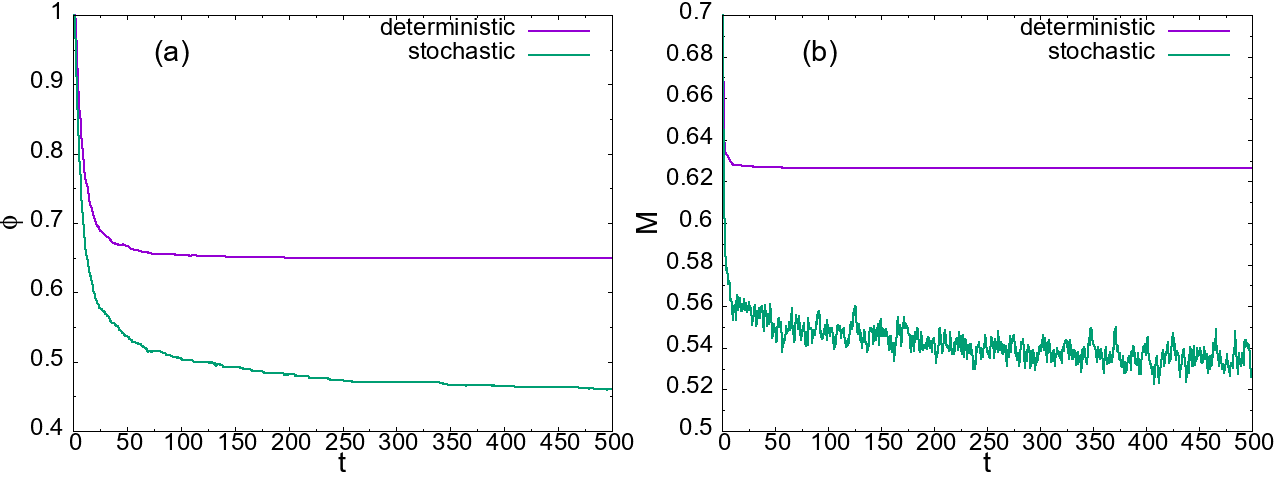}
 \caption{(a) Average fraction of surviving species as a function of time for the deterministic and the stochastic system ($\Gamma=0, \sigma^2=1, N=50, V=200, \mu=-1$, simulations are averaged over $20$ realisations). (b) Time evolution of the average abundance per species for the same parameters as in (a).}
 \label{fig:phi_m}
\end{figure*}

While the deterministic system settles down to a fixed point in the stable regime, the dynamics in finite populations remains subject to fluctuations. As a consequence the mean number of surviving species in the stochastic system (averaged over realisations) shows a mild decay over time, see Fig.~\ref{fig:phi_m}(a). There is therefore no true stationary state for the system with intrinsic noise [one can expect all species eventually to become extinct at extremely long times (scaling exponentially in $V$)]. For the purposes of our analysis we do however have to choose a time at which we study the stochastic system. We choose this time scale ($t\approx 100$) such that the deterministic system has reached its fixed point. We have also verified that results for the stochastic system do not change qualitatively if a longer time scale is chosen (e.g. $t\approx 200$). Further details will be described below.

\begin{figure*}[t]
 \centering
 \includegraphics[width=0.85\textwidth]{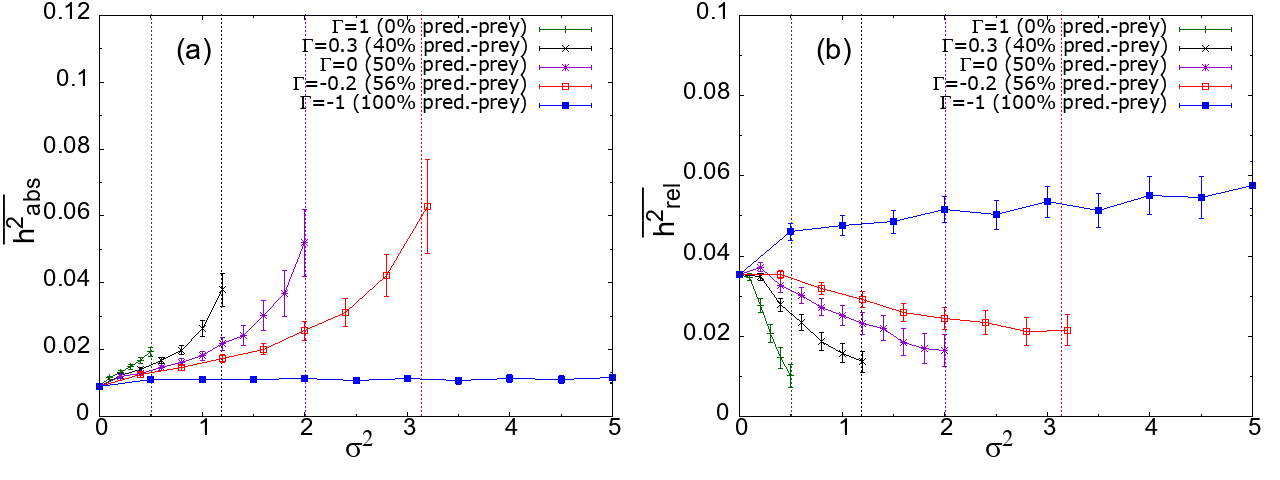}
 \caption{Panel (a) shows the absolute fluctuations $\overline{h_{\rm abs}^2}$ in the stochastic system (averaged over $20$ realisations) as a function of the variance of interaction strength. ($V=100, \mu=-1, N=50$). Panel (b) shows relative fluctuations, $\overline{h_{\rm rel}^2}$. Vertical lines mark the onset of the linear instability in the deterministic gLVE system. }
 \label{fig:h}
\end{figure*}

We first study the magnitude of fluctuations more systematically. To this end we define (for a single realisation),
\be
    h_{\rm rel}^{2}=\frac{\frac{1}{N} \sum_{i=1}^{N} \left( \langle x_{i}(t)^{2} \rangle_{t} - \langle x_{i}(t) \rangle^{2}_{t} \right)}{\frac{1}{N} \sum_{i=1}^{N}\langle x_{i}(t)^{2} \rangle_{t}},
    \label{eq:h_rel}
\ee
where $\avg{\cdots}_t$ stands for an average over time after some equilibration period. More precisely we calculate this average in the stochastic system during the time interval from $t=60$ to $t=100$. Thus, $h_{\rm rel}^2$ describes the relative fluctuations of the species abundances in time. We also define

\begin{equation}
    h^{2}_{\rm abs}=\frac{1}{N} \sum_{i=1}^{N} \left( \langle x_{i}(t)^{2} \rangle_{t} - \langle x_{i}(t) \rangle^{2}_{t} \right),
    \label{eq:h_abs}
\end{equation}
As seen in Fig.~\ref{fig:h}(a) the absolute fluctuations of species abundances increase with $\sigma^2$ for fixed $\Gamma>-1$, showing a particularly steep rise as the linear instability in the deterministic model is approached. One possible reason for this is the growth of the species abundances when the deterministic system approaches the instability [as shown in Fig.~\ref{fig:community_phi_M}(b)]. For $\Gamma=-1$ there is no instability and the predator-prey relations in the ecosystem prevents large abundances. As a result, the increase in absolute fluctuations is only moderate. (We remark that throughout the paper we often distinguish between the cases $\Gamma=-1$ and $\Gamma>-1$. This is a broad classification, we cannot exclude that the system behaves similarly to the fully anti-correlated case when $\Gamma$ is near $-1$.)

Surprisingly, relative fluctuations decrease with $\sigma^2$ for $\Gamma>-1$ [see Fig.~\ref{fig:h}(b)]. The relative effect of the intrinsic noise on the abundances is hence reduced when increasing the diversity in the interactions. However, this effect becomes less pronounced when interactions are anti-correlated (more negative values of $\Gamma$). Indeed, if all interactions are of a predator-prey type ($\Gamma=-1$) relative fluctuations around found to increase only mildly with $\sigma^2$.  

\medskip

\subsection{Community properties}
\begin{figure*}[t]
\centering
 \includegraphics[width=0.85\textwidth]{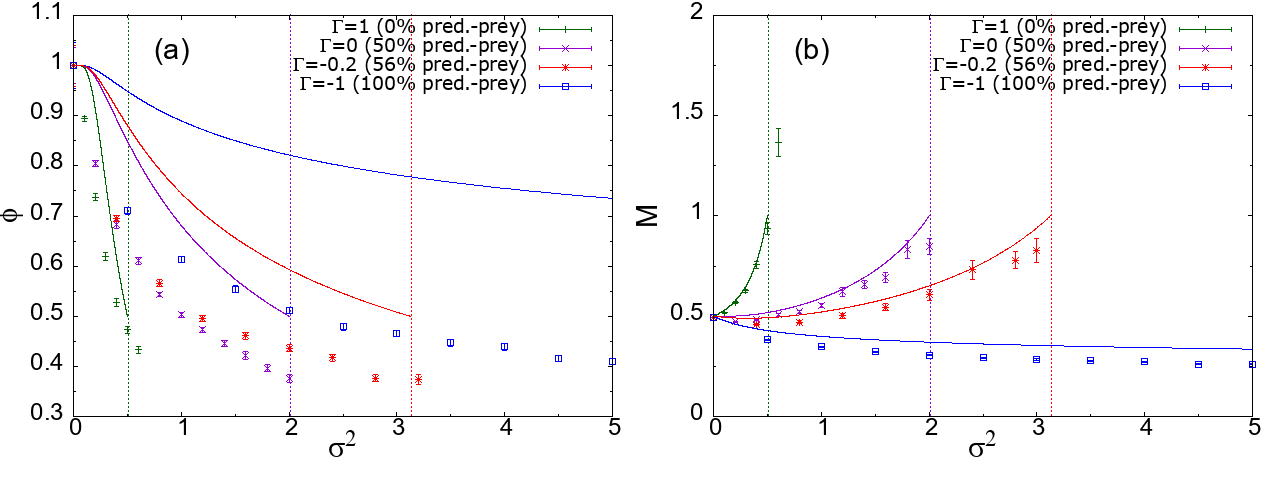}
\caption[Community properties of the stochastic model.]{(a) Fraction of surviving species as a function of the variance of the interaction strengths.  (b) Mean abundance per species as a function of the variance of the interaction strengths. Markers are from stochastic simulations using $N=50$, $V=100$, $\mu=-1$ and averaging over $50$ runs (including error bars), for different values of the $\Gamma$. Solid and dashed vertical lines come from theoretical predictions of the deterministic model (vertical lines are the onset of linear instability).}
\label{fig:community_phi_M}
\end{figure*}

\medskip

In Fig.~\ref{fig:community_phi_M} we study how the size of the surviving community (indicated by the fraction of survivors, $\phi$) and the mean abundance per species, $M$, vary with $\sigma^{2}$, for different values of the correlation parameter $\Gamma$. We note again that these quantities vary in time in the stochastic model (e.g. $\phi$ decreases with time). The results in the figure are therefore only valid at the time they were measured ($t=100$). We have however checked that the qualitative behaviour as a function of $\sigma^2$ and $\Gamma$ does not change at later times.  

Unsurprisingly, the fraction of survivors is systematically lower in the stochastic model (markers) than in the deterministic one (solid lines). This is because fluctuations cause some species to eventually become extinct in the stochastic model while the same species survive in the absence of demographic noise. The average abundance $M$ is measured across all species, including those that go extinct. As a consequence $M$ is also lower in the stochastic model than in the deterministic gLVE.

The data for the fraction of surviving species in Fig.~\ref{fig:community_phi_M}(a) indicates that  the difference between the communities resulting from the stochastic model  and that of the deterministic dynamics becomes more pronounced as the fraction of predator-prey pairs in the interactions increases (i.e., as $\Gamma$ is lowered to approach $\Gamma=-1$).  One possible explanation is that, although communities in the deterministic model are more stable for predator-prey interactions (and there are fewer deterministic extinctions), species abundances are generally smaller [Fig.~\ref{fig:community_phi_M}(b)]. Therefore extinctions caused by fluctuations in the stochastic model become more prevalent as $\Gamma$ becomes more negative.

\begin{figure}[t]
\centering
\includegraphics[width=0.6\textwidth]{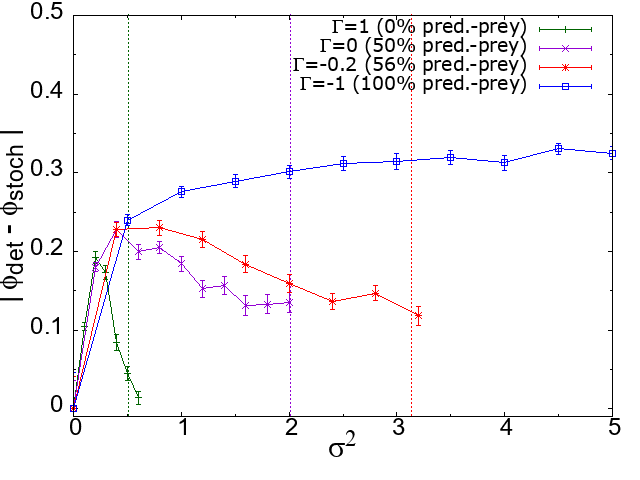}
\caption[Difference between the fraction surviving species in both models.]{Difference between the fraction of surviving species in the deterministic and stochastic models as a function of $\sigma^{2}$, for a system with $N=50$ species, $V=100$, $\mu=-1$ and for several values of $\Gamma$. Vertical lines show the onset of linear instability in the deterministic system.}
\label{fig:phi_difference}
\end{figure}

In Fig.~\ref{fig:phi_difference} we analyse in more detail how the difference between the size of the surviving communities in the two models, measured by $\vert \phi_{\rm det} - \phi_{\rm stoch} \vert$, changes with the complexity of the interaction strengths $\sigma^{2}$.  We observe virtually no difference in the number of survivors in the absence of interactions ($\sigma^2=0$). At the fixed point of the deterministic system all species abundances then take the value $x_i=1$, and extinctions in the stochastic model are  rare. As $\sigma^2$ is increased from zero we observe an initial increase of $\vert \phi_{\rm det} - \phi_{\rm stoch} \vert$ with $\sigma^2$ for all values of the correlation parameter $\Gamma$ that we have tested.

For fully anti-correlated interactions, $\Gamma=-1$, this increase continues to larger values of $\sigma^2$. In this situation, the relative magnitude of fluctuations of abundances in the stochastic model grows with $\sigma^{2}$ [Fig.~\ref{fig:h}(b)]. This leads to an increased number of extinctions in the stochastic model, caused by intrinsic noise.

When interactions are not fully anti-correlated ($\Gamma >-1$), the difference $\vert \phi_{\rm det} - \phi_{\rm stoch} \vert$ becomes smaller as $\sigma^{2}$ increases (except for the region near $\sigma^2\approx 0$ discussed above). Hence the deterministic and stochastic models become more similar to one another, and extinctions occur mostly due to interactions between species (as opposed to demographic noise). This is in line with our earlier observation that relative fluctuations become weaker with increasing $\sigma^2$ when $\Gamma>-1$ [Fig.~\ref{fig:h}(b)], and that the mean abundance per species $M$ increases [Fig.~\ref{fig:community_phi_M}(b)]. Extinctions due to fluctuations hence become more rare with increased variance of the interactions.

\subsection{Species abundance distribution}
\begin{figure}[t]
\centering
\includegraphics[width=0.85\textwidth,right]{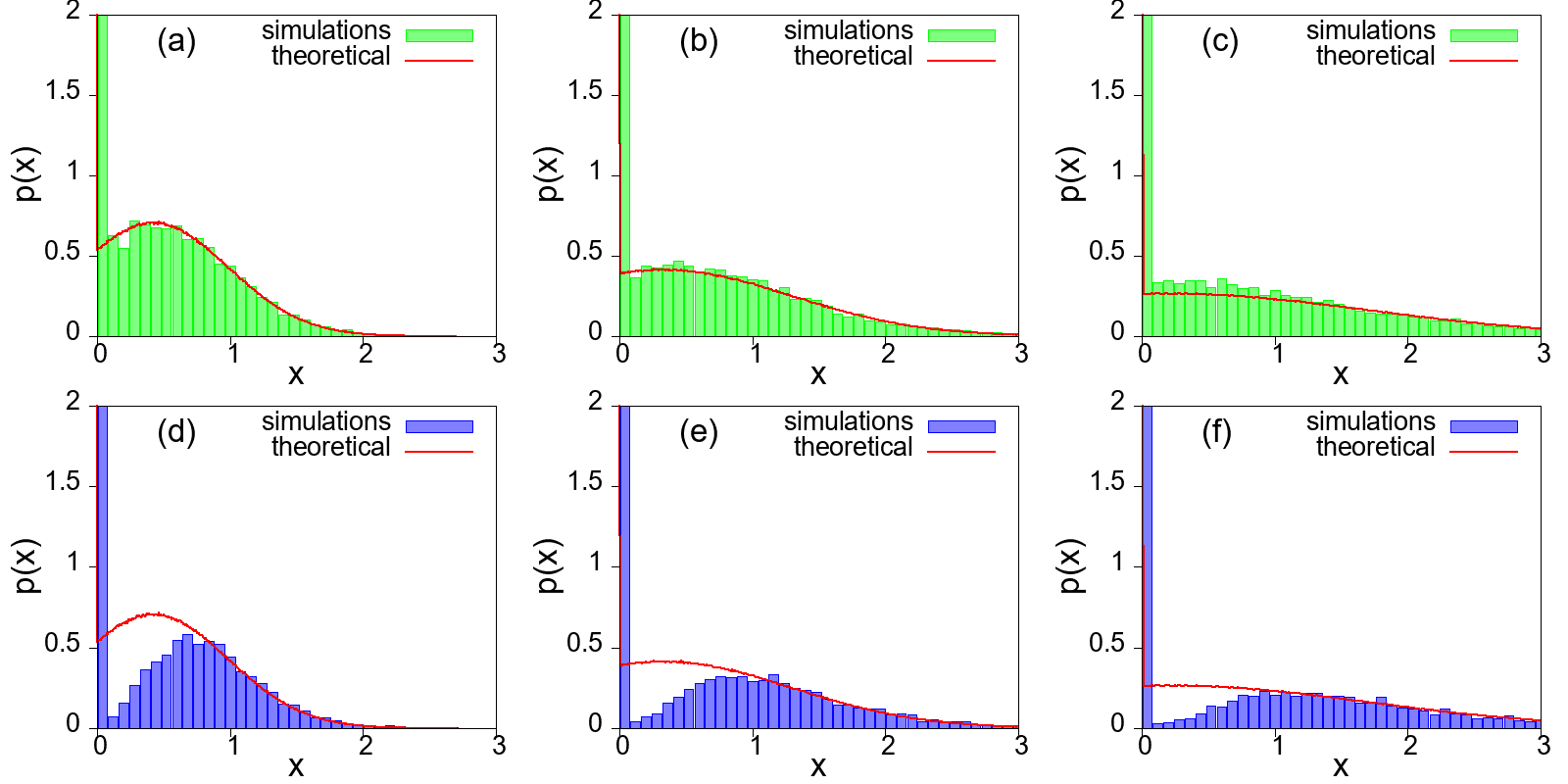}
\caption[Stochastic species abundance distribution for $\Gamma=-0.3$.]{Species abundance distribution for the deterministic model (green top panels) and for the stochastic model (blue bottom panels). The red solid lines (in all panels) are the frequencies of species abundances from theoretical predictions for the deterministic model. The histograms are from numerical simulations, the deterministic model is shown in green [panels (a)-(c)], the individual-based model in blue [panels (d)-(f)]. Parameters are: $N=50$, $V=100$, $\Gamma=-0.3$ and $\mu=-1$, data is from $200$ runs of the interaction matrix. Panels (a) and (d) $\sigma^{2}=1$, (b) and (e) $\sigma^{2}=2$, (c) and (f) $\sigma^{2}=3$.}
\label{fig:sad}
\end{figure}
We next look at the effects of intrinsic noise on species abundance distributions (SAD).  In deterministic model the mean abundance increases with $\sigma^2$, and the SAD becomes wider. Species with low abundance in the deterministic model are generally more likely to go extinct in the face of noise than those with high abundance. So we expect that any deviation of the SAD in the stochastic model from that of the determinisitic model will result mostly at small abundances $x$. This is confirmed in Fig.~\ref{fig:sad}.

Broadly, the picture is here that the gap between the SADs of the deterministic and stochastic models is localised mostly at $x<x_c$, with $x_c$ some critical abundance, and that for abundances $x>x_c$ the SAD is mostly unaffected by intrinsic fluctuations. (This is not a strict classification, we are not suggesting that the SADs of the two models are identical for $x>x_c$. But we think the notion of a value of $x$ below which most of the deviations are seen is useful.) The numerical value of $x_c$ will depend on the model parameters. In particular, the data in Fig.~\ref{fig:sad} suggests that $x_c$ grows with $\sigma^2$. This is because absolute fluctuations of species abundances become larger with $\sigma^2$, as previously shown in Fig.~\ref{fig:h}(a). Consequently, as $\sigma^2$ is increased, extinctions due to demographic noise can occur for species with higher abundances.

\begin{figure*}[t]
 
\includegraphics[width=0.85\textwidth,right]{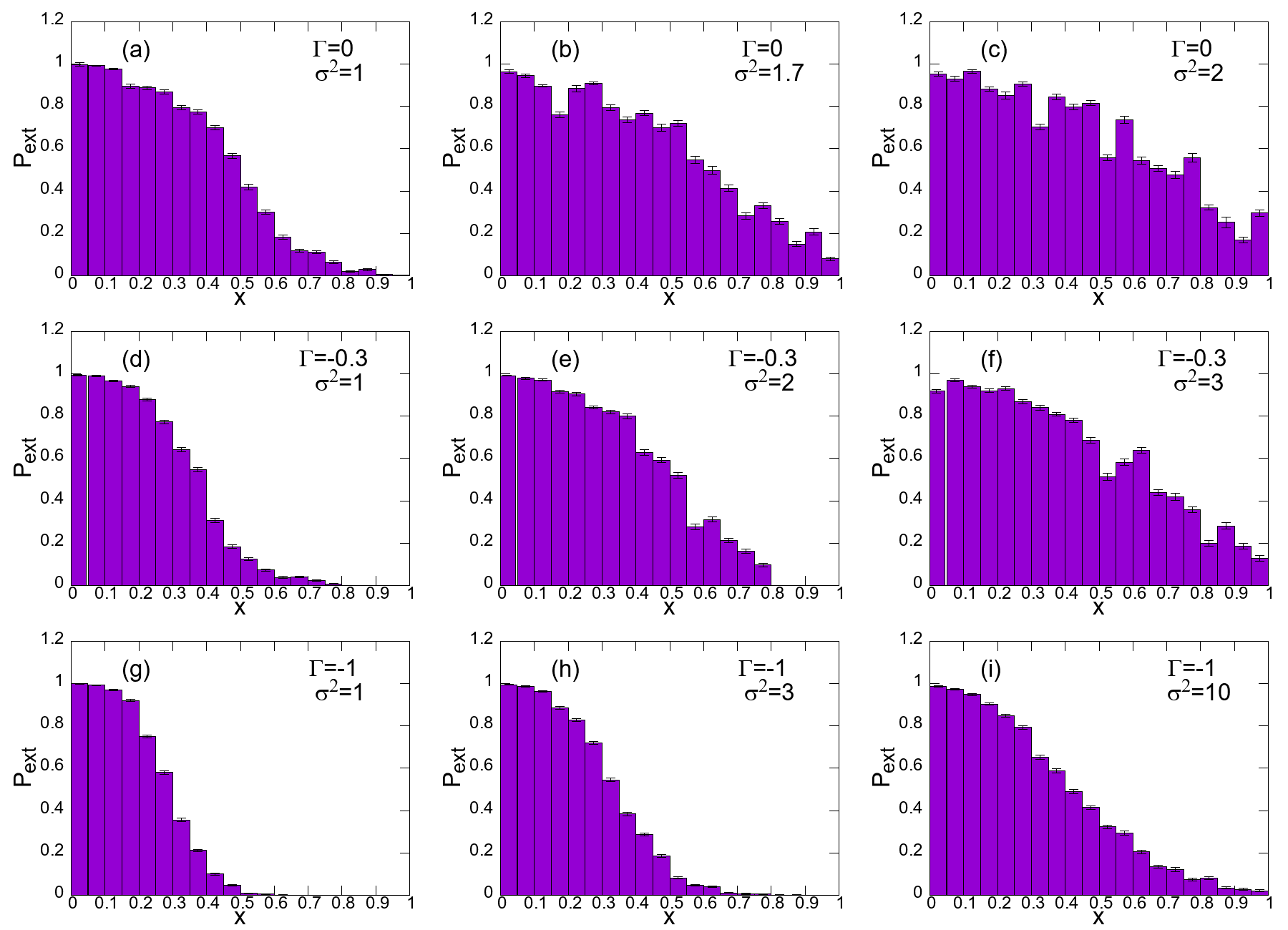}
 \caption{Probability for a species to go extinct in the stochastic model as a function of its abundance at the deterministic fixed point. Data is from simulations ($N=50$, $V=100$, $\mu=-1$, $30$ realisations of the interaction matrix, each with $50$ realisations of the stochastic dynamics), and shown for different combinations of the correlation parameter $\Gamma$ and the variance of interactions, $\sigma^2$.}
\label{fig:p_ext}
\end{figure*}

In Fig.~\ref{fig:p_ext}, we show the probability for a species to go extinct if it has attained an abundance $x$ in the deterministic model. To measure this, we first fixed an interaction matrix, integrated the gLVE numerically, and determined the abundances of the surviving species in the resulting equilibrium. We then ran multiple simulations of the stochastic model, with the same interaction matrix, and determined how likely the different species were to survive up to time $t=100$. This was then repeated for different realisations of the interactions, for a given choice of $\sigma^2$ and $\Gamma$. We observe in Fig.~\ref{fig:p_ext} that species with higher abundance in the deterministic model also become prone to extinction as $\sigma^{2}$ increases. Further, the data demonstrates that making the interactions more anti-correlated (increasingly negative values of the correlation parameter $\Gamma$), the probability to go extinct for species with high to moderate $x$ decreases [at fixed $\sigma^2$, compare panels (a), (d), (g) in Fig.~\ref{fig:p_ext}]. Indeed, reducing $\Gamma$ at fixed $\sigma^{2}$ reduces absolute fluctuations as previously shown in Fig.~\ref{fig:h}(a).

We note that the extinction probabilities in Fig.~\ref{fig:p_ext} are, in isolation, not immediately a statement about the total number of extinctions in the system due to fluctuations. What is shown  in Fig.~\ref{fig:p_ext} is the probability that a species becomes extinct in the stochastic system provided it attains a given abundance in the deterministic system. The total number of extinctions due to noise results from a combination of these probabilities and the abundance distribution in the deterministic system. For $\Gamma>-1$, these abundances move towards higher values as the variance of interactions, $\sigma^{2}$, is increased [see Fig.~\ref{fig:community_phi_M}(b)]. There are therefore two competing effects acting on total extinctions as $\sigma^2$ is increased, increasing abundances (reducing the propensity for extinction) and the increasing probability to go extinct for any fixed deterministic abundance, $x$.

\subsection{Abundance dispersion}
As a further point we now study of the differences between the abundances a species attains in the stochastic and the deterministic systems with the same interaction matrix. Generally we find that the abundances in the stochastic system, although fluctuating in time, are closely related to the deterministic fixed point abundances for systems that are far away from any instability.  However, as we will discuss, this is not the case when parameters are such that the deterministic system approaches the unstable phase.

\begin{figure*}[t]
\centering
\includegraphics[width=0.85\textwidth,right]{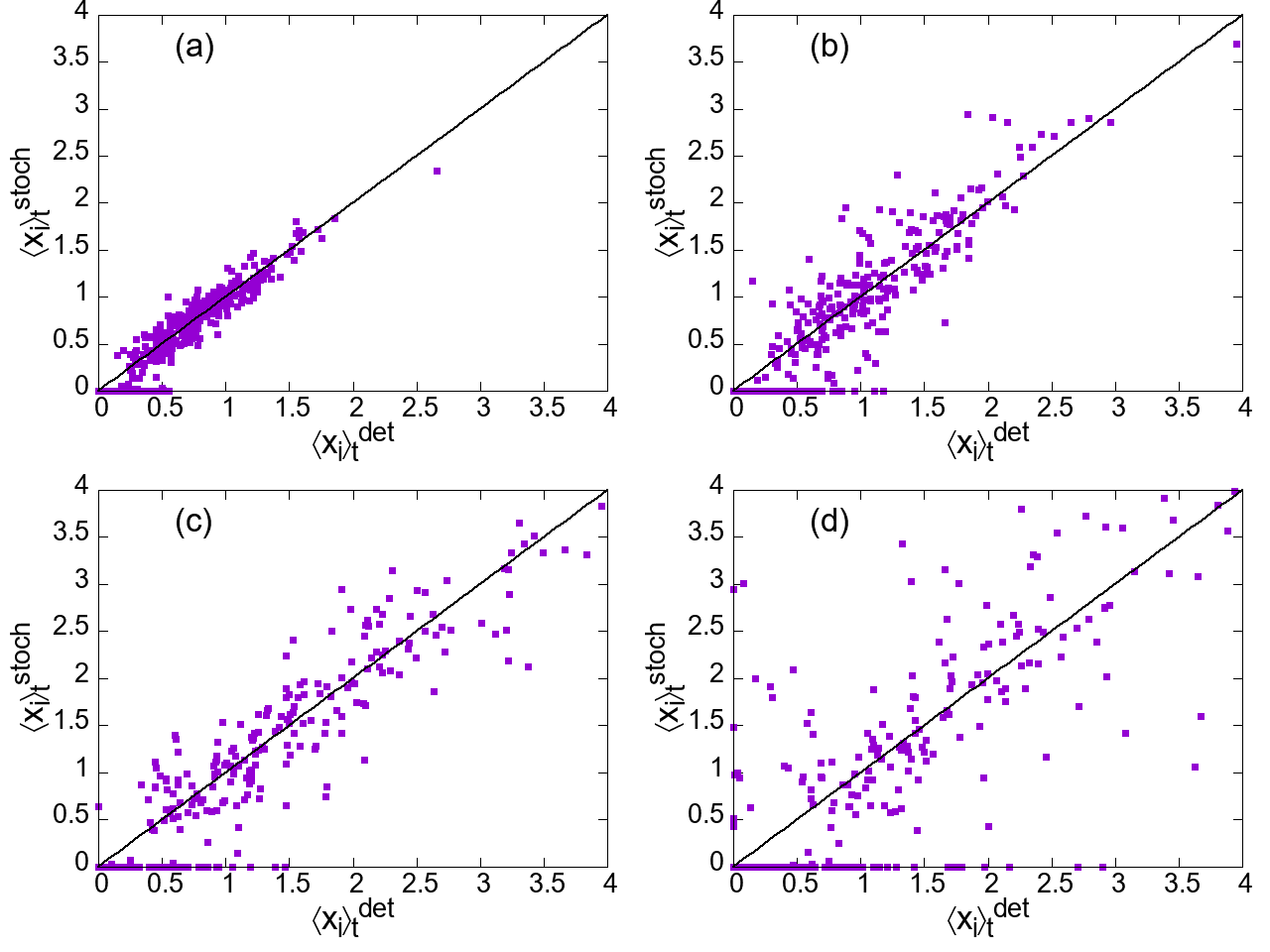}
\caption[Scattered plot of the deterministic and stochastic abundances for $\Gamma=-0.3$.]{Dispersion between the temporal mean value of species abundances for the deterministic and stochastic models. $N=50$, $V=100$, $\Gamma=-0.3$ and $\mu=-1$, using $10$ different interaction matrices. (a) $\sigma^{2}=1$. (b) $\sigma^{2}=2$. (c) $\sigma^{2}=3$. (d) $\sigma^{2}=3.5$.  (We remark that the density of points in the plots reduces as we increase $\sigma^2$. This is because the species abundances become larger and because we fix the range of abundances in the different panels for better comparison.)}
\label{fig:dispersion}
\end{figure*}

To study how abundances in the two different models differ from one another,  we  produce scatter plots of these two quantities. Each point on the graph is a species, and the coordinates of the points are the species' abundances in the deterministic and the stochastic systems with the same interaction matrix. If the abundances in both models are fairly similar, one would expect to see a dense region of points near the diagonal. Conversely, points away from the diagonal indicate differences between the abundances in the stochastic and deterministic models. We use the temporal average value of the abundances measured in the steady state of the deterministic system. We average the abundances in the stochastic system from $t = 60$ to $t = 100$. For the deterministic abundances, we average from $t = 90$ to $t = 100$, this is sufficient to capture the deterministic fixed point of the dynamics. For all simulations we use $N = 50$, $V = 100$, and $\mu = -1 $. Data is shown in Fig.~\ref{fig:dispersion}. Each of the scatter plots in the different panels is from an aggregate of data from $10$ different interaction matrices.

The figure demonstrates that the abundances of the two models are quite similar for small $\sigma^{2}$, but differ more substantially when approaching the unstable regime, therefore appearing more dispersed in the graph. While we only show data for one particular choice of the correlation parameter $\Gamma$, we have tested other values as well, and found similar behaviour. 

To capture these findings more compactly we define the average absolute difference between abundances as
\begin{equation}\label{eq:diff_x}
\delta x=\frac{1}{N_S}\sum_{i\in{\cal S}} \left|\langle x_{i} \rangle_{t}^{\rm det} - \langle x_{i} \rangle_{t}^{\rm stoch}\right|,
\end{equation}
 where we restrict the sum over $i$ to species that survive both in the deterministic and in the stochastic model (as indicated by the set ${\cal S}$). Numerically, we identify these as species with $\langle x_{i} \rangle_{t}^{\rm det}>0.001$ and $\langle x_{i} \rangle_{t}^{\rm stoch}>0$. The normalising factor $N_S$ in Eq.~(\ref{eq:diff_x}) is the number of such surviving species. Including only the set of survivors avoids that the outcome is dominated by species that go extinct in the stochastic model, but survive in the absence of noise.  
 
\begin{figure}[t]
\centering
\includegraphics[width=0.6\textwidth]{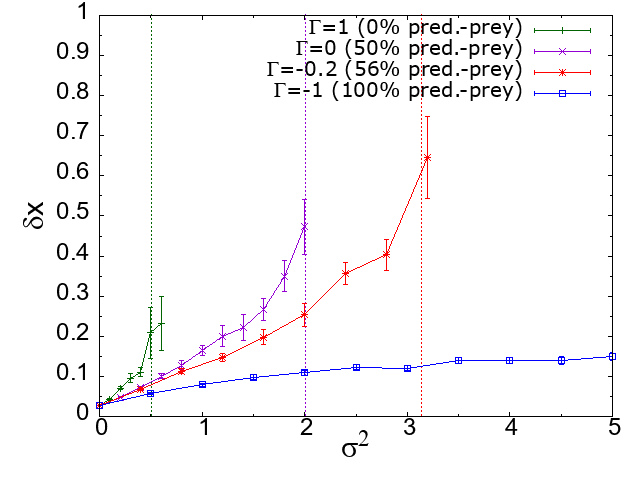}
\caption{Average absolute difference between abundances of surviving species in the deterministic and stochastic models. Parameters are $N=50$, $V=100$, $\mu=-1$, data is from an average over $20$ realisations of the interaction matrix. Vertical lines show the onset of linear instability in the deterministic system.}
\label{fig:delta_x}
\end{figure}

Indeed, as shown in Fig.~\ref{fig:delta_x}, $\delta x$ grows with $\sigma^{2}$, with quicker increase for positive $\Gamma$ and only slowly for systems with anti-correlated interactions ($\Gamma \approx -1$), in-line with the behaviour of the absolute fluctuations of abundances shown in Fig.~\ref{fig:h} (a). Similar behaviour is found if all species are included.

\subsection{Summary of stochastic effects}
We find that the fraction of surviving species, $\phi$, is systematically lower in the individual-based model than in the deterministic gLVE. This is not surprising, and mainly serves to confirm that additional extinctions occur in the stochastic model, driven by demographic noise. The mean abundance per species, $M$, is also found to be lower in the stochastic model. 

We also find, that at a fixed value of the correlation parameter $\Gamma$, absolute fluctuations of species abundances in the individual-based model grow with the variance $\sigma^2$ of the interaction coefficients. Hence we observe a more dispersed pattern in scatter plots of abundances in the stochastic models versus those in the absence of noise. Differences between the species abundance distributions of the deterministic and stochastic models  extend to larger and larger abundances as the complexity of interactions is increased and absolute fluctuations grow. Species with higher abundances in the deterministic model hence become more vulnerable to fluctuations when the complexity of interactions is high. 

It is known that abundances in deterministic model grow with the complexity $\sigma^2$. This growth, combined with that of absolute fluctuations, creates competing forces on the relative magnitude of abundance fluctuations. We find that relative fluctuations decrease as the interactions become more varied when the system is not primarily composed of predator-prey pairs ($\Gamma > -1$).  The total number of extinctions due to noise then reduces with $\sigma^2$, despite the fact that species with higher deterministic abundances become affected by noise. This is because the increase of abundances in the deterministic model with $\sigma^2$ more than compensates for the increase in absolute fluctuations.

For systems composed only of predator-prey pairs ($\Gamma = -1$), relative abundance fluctuations increase with complexity. The number of noise-driven extinctions then also grows. The difference between the communities in the stochastic and determinstic systems becomes larger as the proportion of predator-prey pairs increases. Systems with $\Gamma=-1$ are more stable deterministically, and overall there are fewer extinctions, but there are also more species with lower abundances, hence the community of species is more vulnerable to demographic noise.

\section{Conclusions and discussion}\label{sec:concl}

In summary we have used an individual-based model to study the effects of demographic noise in Lotka--Volterra systems with complex random interactions. As we have shown, intrinsic stochasticity leads to additional extinctions, and changes the composition of the surviving community. Species which have a low abundance in the absence of noise are particularly prone to the effects of fluctuations. How strong these effects are depends on the detailed circumstances, in particular on the variance of the interactions and on the number of predator-prey relations in the interaction matrix.

Our work complements existing literature approaching complex ecological communities with deterministic differential equations (with random, but fixed interaction coefficients), and a large body of work assessing stability of equilibria based on the spectra of random matrices. Neither of these approaches explicitly account for granular populations of individuals, subject to discrete birth and death events. The effects of the resulting demographic noise are particularly relevant when the number of individuals per species is small enough that abundances cannot be approximated by continuous variables. This may be relevant for example in the case of microbial communities or for models in game theory, where effects of intrinsic noise are known to be able to significantly change the outcome.

There are several limitations to our work. First, we focus on parameter choices for which the deterministic gLVE system has a unique stable fixed point. In parameter regimes where abundances diverge in the deterministic system the number of Gillespie steps required per unit of simulated time of the individual-based model also diverges. This makes simulations up to a sensible time horizon difficult. At the same time we expect that the relative effect of intrinsic fluctuations is then also rather limited, similar to what we have seen in Fig.~\ref{fig:h}(b). One aspect one could consider in future work is to characterise the effects of stochasticity just above the linear instability lines in Fig.~\ref{fig:pg}, when the divergence of abundances has not yet set in \cite{bunin2016interaction, bunin2017}.

Secondly, our study is based on simulations, no analytical theory is available at present to systematically study extinctions due to intrinsic stochasticity. Finally, we restrict our analysis to relatively simple interaction matrices, with no particular sub-structure. There are therefore multiple avenues for future work. This can include expansions of the Kramers--Moyal or van-Kampen type (in the limit of large, but finite $V$), which would lead to sets of stochastic differential equations which capture the two types of randomness -- that of the interaction matrix, and intrinsic noise in finite populations. The latter would be approximated as dynamic Gaussian in those series expansions. Initial steps have been taken in \cite{stochastic1}. It would be interesting to see if further theoretical analysis and characterisation of the intrinsic fluctuations is possible, potentially making further simplifications such as the linear-noise approximation. One could also attempt to connect such approaches with existing replica analyses at non-zero temperature \cite{altieri2021}. Finally, our work can be extended to more intricate interaction structures (e.g. those in \cite{allesinatang2}). The role of competition, mutualism, and nestedness could be studied, and it would also be interesting to see what the effects of intrinsic noise are in complex ecosystems with spatial structure. Further work could also focus on the possibility of noise-induced patterns in individual-based models of spatial complex ecosystems \cite{gravel, baron2020dispersal}.

\section*{Acknowledgements}
We acknowledge funding from the Agencia Estatal de Investigaci\'on (AEI, MCI, Spain) and Fondo Europeo de Desarrollo Regional (FEDER, UE) under Project PACSS (RTI2018-093732-B-C21/C22), and the Mar\'ia de Maeztu Program for units of Excellence in R\&D, Grant No.~MDM-2017-0711 funded by MCIN/AEI/10.13039/501100011033. This work was partially supported by MINEICO (Spain) and Agencia Estatal de Investigaci\'on (AEI) with grant number PID2019-106811GB-C33 (funded by AEI/10.13039/501100011033; FL).

\section*{References}
\providecommand{\newblock}{}


\end{document}